# Degenerate phase-matching for multi-wavelength nonlinear mixing in aperiodic lattice lasers

WEI JIANG, LI HUA, SUBHASISH CHAKRABORTY*

*Department of Electrical and Electronic Engineering, University of Manchester, M13 9PL, UK*
*s.chakraborty@manchester.ac.uk*

**Holographically-designed aperiodic lattices have proven to be an exciting engineering technique for achieving electrically switchable single- or multi-frequency emissions in terahertz (THz) semiconductor lasers. Here, we employ the nonlinear transfer matrix modeling method to investigate multi-wavelength nonlinear (sum- or difference-) frequency generation within an integrated THz (idler) laser cavity that also supports optical (pump and signal) waves. The laser cavity includes an aperiodic lattice, which engineers the idler photon lifetimes and effective refractive indices. The key findings are: (i) the nonlinear conversion efficiency reveals resonant enhancement at those idler frequencies where the photon lifetime is high; (ii) the resonant phase-matching process between the pump and idler waves has a one-to-one link with the engineered effective index dispersion; (iii) in absence of any other dispersion, the lowest threshold, multi-wavelength defect modes of the aperiodic lattice laser have degenerate phase-matched pump frequencies. This set of results will potentially have a significant impact on the wavelength multiplexing in electronically switchable THz-over-fiber communication systems [1].**

Nonlinear optical frequency conversion, such as intra-cavity sum- or difference-frequency generation (SFG and DFG), is a well-established technique, which enables, for example, frequency mixing between near-infrared (NIR) and terahertz (THz) waves in a THz quantum cascade laser (QCL) cavity. This has been exploited to send THz signals down an optical fiber as optical sidebands, where standard telecom wavelengths, operating at either 1.3 μm or 1.54 μm, are used as carriers – these are beneficial for wavelength-division multiplexing (WDM) applications. This technology was first demonstrated in Refs. [2-4], using a compact electrically driven QCL, based on GaAs/AlGaAs materials, by exploiting modal polaritonic phase-matching, following which it was adopted and further exemplified both experimentally and theoretically by the present authors' group, who subsequently demonstrated the first coherent detection of THz signals in optical fiber systems [1, 5-6].

In Ref. [4], the opto-THz frequency mixing was made more flexible via the exploitation of the metal-metal geometry because, although the NIR wave refractive index is independent of the THz waveguide geometry, the idler wave refractive index is strongly affected by the laser ridge width and so, the mixing process, including phase-matching (PM), could be tuned across devices with very different waveguide widths. In this paper, we establish, using computer simulations, that a holographically-designed, multi-wavelength aperiodic lattice (AL) laser could, via group index ($\tilde{n}_g$) and effective index ($\tilde{n}_{eff}$) dispersion engineering of the idler (THz) waves, introduce additional powerful control over both the nonlinear frequency mixing and the PM process: we unequivocally demonstrate that the nonlinear conversion efficiency ($\eta$) exhibits resonant enhancement which resembles the $\tilde{n}_g$ (or the photon lifetime) modulation, whereas the resonant PM process between the pump, signal and idler waves has a one-to-one link with the $\tilde{n}_{eff}$ dispersion. We also observe that in absence of any other dispersion, the degenerate defect modes of the AL laser, although distinctly separate, have near identical (i.e. degenerate) phase-matched pump frequencies. As illustrated in Fig. 1, practical execution of this scheme could lead to electronically switchable multi-wavelength optical sidebands for THz WDM communications using a single integrated device [1]. More fundamentally, this work lays the theoretical foundation of building such systems, be it classical or quantum, based on holographically-designed AL lasers. In a classical architecture, a THz wavelength division multiplexed passive optical network could be envisaged [7], enhancing the spectral efficiency by adopting the wavelength-to-the-user approach (each wavelength supporting a data bandwidth of several tens of GHz), whereas, for a quantum communication system, parallelization of quantum key distribution [8], for example, by THz sub-carrier multiplexing, could also be realized, where multiple closely-packed sub-carrier channels can be independently used, providing high spectral efficiency and sending several parallel keys. While chip-based optical frequency combs can potentially replace large numbers of lasers with the individual comb lines [9], their use in WDM systems have so far been limited and the high costs of WDM components and operating expense has also been an issue. Transferring the multi-wavelength AL control function $\rho(f)$, designed to carry a user-defined RF channel separation $\Delta f_{RF}$ across the THz-QCL frequency and then onto NIR optical fields, as shown in Fig. 1, provides an additional powerful means towards THz WDM communications.

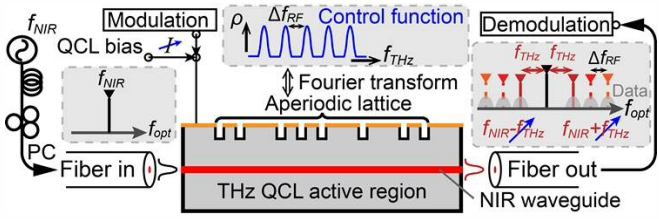

Fig. 1. Multi-wavelength THz WDM architecture, reproduced from Refs. [1, 5, 6]. The AL offers a user-defined control function $\rho(f)$ at the THz band, characterized by the RF channel separation $\Delta f_{RF}$; varying the QCL bias leads to selection of electronically switchable THz (idler) mode $f_{THz}$. The pump laser ($f_{NIR}$), mixed with $f_{THz}$, generates the SFG and DFG optical signals, each of which mimics the idler response and $\Delta f_{RF}$.

The notion of utilising the dispersion of a photonic lattice is fairly well-known [10, 11] – a carefully-designed lattice could tailor the dispersion relation via $\tilde{n}_{eff}$ engineering, so that a phase mismatch due to normal dispersion of the material is compensated in a given triplet of waves by the lattice momentum [12]. Additionally, a useful property of photonic lattices is the emergence of spectral transmission resonances that appear close to the band gaps or at the frequency of the defect level, which is accompanied by an associated increase in the photon lifetime or the density-of-modes (DOM) and corresponds to a slowing down of the wave (i.e. a resonantly boosted $\tilde{n}_g$). Consequently, enhancement of an SFG or DFG as a result of resonances for a uniform or single-defect lattice has been widely investigated [13, 14]; however, 'multi-defect' or 'aperiodic' lattices have not attracted much attention. To this end, it has been demonstrated that a holographically-designed AL, offering a user-defined multiband control function $\rho(f)$, post-processed using Focused Ion Beam etching into an off-the-shelf Fabry-Perot (FP) QCL cavity, could select a number of modes either from a single band or simultaneously from multiple bands within the available gain bandwidth of the laser and then could also switch the emission frequency, with purely electrical control [15-22]. At the heart of this control lies the photonic DOM engineering mentioned above, which in this case has a close correlation with $\rho(f)$ and modulates the threshold gain spectral profile for the FP-QCL device, achieving an amplified reflectivity ($R$) response, which tends to approach infinity at the high-DOM frequencies when the material gain ($g$) reaches the threshold – these infinite reflectivity points, plotted in the frequency-gain ($f,g$) plane (Fig. 2(d1)-(d3)), are singularities [19].

It has been demonstrated experimentally that integrating the AL THz QCL technology with the intra-cavity SFG/DFG generation process leads to electronically switchable multi-wavelength optical sidebands, i.e. the idler wave AL filter modulation can be encoded to the signal band via nonlinear frequency conversion [1]. In such a multi-wavelength nonlinear mixing system, the PM process of the three waves, i.e. idler (THz), pump and signal waves, is highly sensitive to their respective refractive indices [5] and the photonic DOM control, exerted by $\rho(f)$, engineers the dispersion of an otherwise unperturbed FP cavity. However, what has been missing so far, is a comprehensive explanation of the underlying theoretical mechanisms of the entire process. Here, we address this by employing the nonlinear transfer matrix method (TMM). Specifically, we investigate the nonlinear frequency mixing within four different distributed feedback (DFB) grating designs, where the corresponding $\rho(f)$, exclusively functioning for the idler waves, ranges between a uniform (single band), a single-defect (two bands) and two holographically-designed, aperiodically distributed multi-defect (three and multiple bands, respectively) structures. For comparison, the first three gratings (Fig. 2(a1)-(a3)) are of identical length; the fourth lattice (double the length, Fig. S2(a)) is discussed in Supplement 1. These geometries were investigated previously in great detail, both experimentally and theoretically, but just for the THz-QCL mode control [19, 21]. In our simulation, as SFG and DFG produce identical plots, only SFG is considered, i.e. $f_s = f_p + f_i$, where both energy and wave-vector conservation need to be guaranteed for a successful nonlinear mixing process, i.e. $n_s f_s = n_p f_p + n_i f_i$ [2]. Calculation methodology, including the TMM framework, is described in Supplement 1.

The nonlinear TMM results (see parameters and other details in caption of Fig. 2) within the idler frequency ($f_i$) band reveal that the high DOM, expressed here in terms of group delay $\tau_g$ or group index $\tilde{n}_g$ (Fig. 2(b1)-(b3)), leads to the resonant enhancement of $\eta$ (the color 3D plots, Fig. 2(e1)-(e3)). Distinct resonances are observed for the lowest threshold band-edge modes for the uniform grating (Fig. 2(d1)), whereas for the single-defect and the aperiodic DFB gratings, the nonlinear process is more significant for the lowest threshold defect modes (Fig. 2(d2) and (d3), respectively), revealing strong $\eta$ resonances. While it is clear that all $\eta$ resonances fundamentally originate from their respective resonantly enhanced DOM, for additional understanding, the color 3D ($\eta$) plot for each structure is decomposed as follows: first, $\eta$ is plotted within the $f_i$ band alone (Fig. 2(f1-f3)) where, to maximize the respective resonances, the pump frequencies are tuned to their corresponding phase-matched values, and for each case, the highest $\eta$ is achieved at the low threshold idler ($f_i$) modes of frequencies ($f_i \equiv$) $f_x$ ($x = 1$ or 2 for the uniform DFB; $x = 1, 2$, or 3 for the single-defect DFB; $x = 1, 2, 3, 4, ...$ for the aperiodic DFBs), making the strong correlation between $\eta$ and DOM (e.g. $\tilde{n}_g$) obvious.

Next, the efficiency curves ($\eta$ vs. $f_p$) for the lowest threshold modes $f_x$, each showing traditional sinc- function-like characteristics, are depicted in the right panels of Fig. 2(e1–e3) and Fig. S2(e). The distinct sinc-plots for the band edge modes, visible for the uniform and single-defect cases, confirm that the phase-matched pump frequencies ($f_p$) are dependent on the cavity resonant (idler) frequencies, essentially originating from the strong $\tilde{n}_{eff}$ dispersion around the Bragg frequency (discussed below). This observation is consistent with our previous work, where we experimentally demonstrated multi-mode up-conversion of THz waves inside an FP-QCL cavity – $\eta$ for each up-converted NIR signal has a distinct PM curve, but follows the FP-QCL cavity $\tilde{n}_{eff}$ dispersion [5]. However, here, the crucial new information is revealed from the two defect modes of the short AL ($f_3$ and $f_4$), which are separated by $\sim 0.026 f_B$ [19], and the ten defect modes of the long AL [21], which are separated by an average spacing $\sim 0.01 f_B$, spanning over $\sim$290 GHz, between $f_3$ and $f_4$ (Fig. S2(e)), but both cases still show nearly identical phase-matched $f_p$ values, i.e. almost no adjustment is necessary for the pump source. For comparison, an FP-QCL device (cavity length: 4.6 mm, free-spectral range: $\sim$8.5 GHz) with modes separated by $> 74$ GHz (i.e. $\sim 0.026 f_B$) would have necessitated a phase-matched $f_p$ separation of $\sim$2 THz [5].

Fundamental reasons for these observations can be understood by investigating the phase-matched $f_p$ variation against $f_i$ (Fig. 2(g1–g3) and Fig. S2(g)), which follows a one-to-one correlation with the grating dispersion ($\tilde{n}_{eff}$ vs. $f_i$ plots, Fig. 2(c1–c3) and Fig. S2(c)). As both energy and wave-vector conservation need to be guaranteed, for a successful nonlinear mixing (i.e. the SFG) process, the phase-matched $f_p$ can then be expressed as: $f_p = (\tilde{n}_{eff} - n_s)/(n_s - n_p)f_i$, which indicates that $f_p$ can be managed by suitable engineering of $\tilde{n}_{eff}$ of the DFB cavity and this is what our simulation has successfully demonstrated. Specifically, for the uniform and single-defect DFB gratings, $\tilde{n}_{eff}$ exhibits a strong dependence on $f_i$ between the band edges; for the uniform grating, as $f_i$ changes from $f_1$ = 2.8298 THz to $f_2$ = 2.9864 THz, the corresponding $\tilde{n}_{eff}$ changes (Fig. 2(c1)) from 3.638 to 3.575 (<1% variation across $\overline{n_{eff}}$, with $\kappa L_g$ = 3), leading to a significant (>11%) change in the phase-matched $f_p$ from 215.68 to 192.8 THz, and likewise, for the single-defect case, the different values of $\tilde{n}_{eff}$ for the two band-edge modes ($f_1$ and $f_2$) give rise to different phase-matched $f_p$ of 211.6 and 196.88 THz, respectively.

In contrast, for the short (long) AL, $\tilde{n}_{eff}$ dispersion has been designed to have double (multiple) discontinuities or two (multiple) defect levels within the photonic stop band, which lead to the two (multiple) defect modes. We have shown before that these modes are robust against $\kappa L_g$ variations (and $gL_g$ too, see below) [19, 20], as they are of a similar nature to that of the single-defect grating and $\tilde{n}_{eff}$ of these defect modes is also pinned at $\overline{n_{eff}}$. Consequently, the phase-matched $f_p$ exhibit the nearly degenerate values for the entire ensemble of defect resonances, which are ~204 THz and ~202 THz (Fig. 2(e3) and Fig. 4(a), respectively). The robustness of the AL defect modes can be further highlighted by comparing their $\eta$ resonance positions for $gL_g \neq 0$ (Fig. 3 (a1)–(a3), (b1)-(b3) and Fig. S3), where the effect of gain on the nonlinear TMM simulation is presented – unlike the band edge modes, the phase matched $f_p$ for the defect modes remain robust against the gain variation.

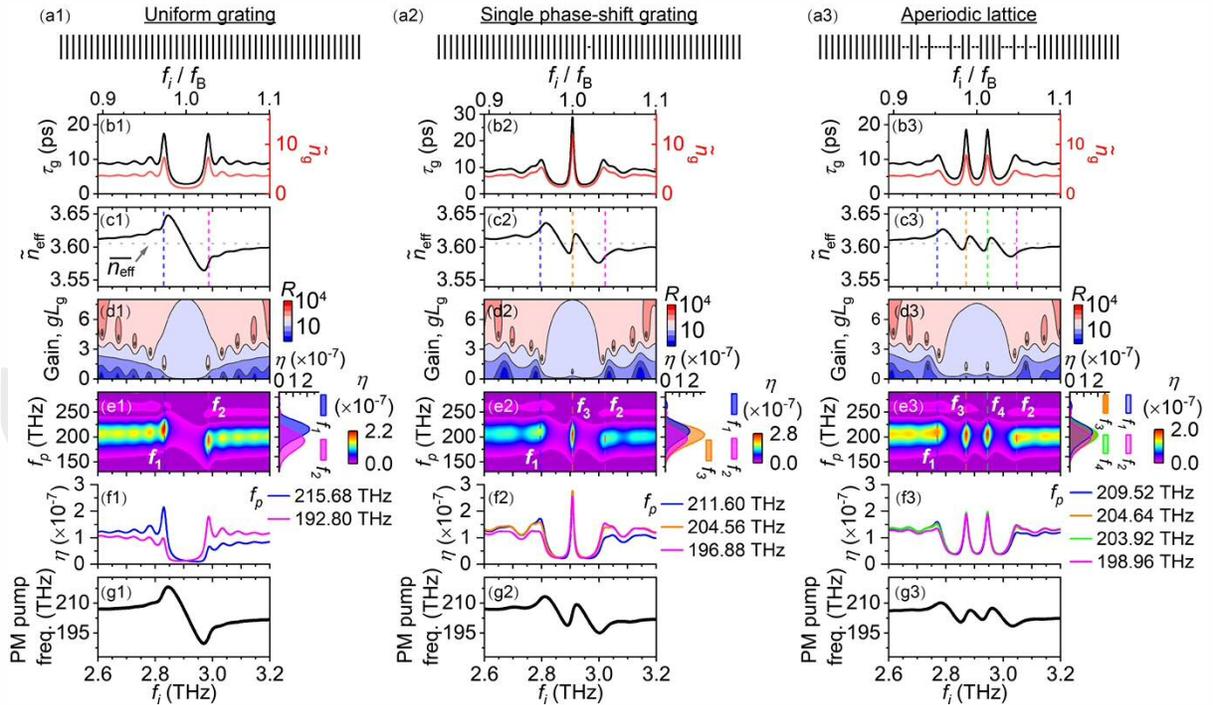

Fig. 2. Nonlinear TMM simulation results as a function of the idler frequency ($f_i$) for the uniform, single phase-shift and aperiodic gratings (columns 1–3 respectively); the resonant responses of group delay ($\tau_g$) or group index ($\tilde{n}_g$) functions lead to the resonant enhancement of the nonlinear conversion efficiency ($\eta$) (for SFG), whereas the effective index ($\tilde{n}_{eff}$) dispersion determines the choice of the necessary phase-matched pump frequency ($f_p$). The grating coupling constant $\kappa = 2\Delta n/\lambda_B$, where $\Delta n$ (refractive index contrast) = $n_1 − n_2$, $L_g$ is the length of the grating, and $\lambda_B$ (Bragg wavelength) = $c/f_B = 2\overline{n_{eff}}\Lambda$ ($\overline{n_{eff}}$ is the average refractive index; $\Lambda$ is the minimum separation between the grating elements; $c$ is the speed of light in free space). Parameters used are: $\overline{n_{eff}}$ = 3.605, $\kappa L_g$ = 3, $gL_g$ = 0, $\Lambda$ = 14.3 μm, $\chi^{(2)}$ (GaAs) = 100 pm/V, the injected pump power $P_{pump}$ = 3 mW and the effective mode overlaps $1/A$ = $4\times10^8$ m$^{-2}$ (discussed in Supplement 1). (a1) – (a3) Grating designs – lines represent low index; horizontal dashes represent defects in multiples of $\Lambda/2$; (b1) – (b3) $\tau_g$ (left) and $\tilde{n}_g$ (right); (c1) – (c3) $\tilde{n}_{eff}$; (d1) – (d3) Reflection gain (power) $R$ over the $f_i$-$gL_g$ plane; (e1) – (e3) $\eta$ over the $f_i$–$f_p$ plane (the color 3D plots) and $\eta$ against $f_p$ (right panel); (f1) – (f3) $\eta$ against $f_i$; (g1) – (g3) Phase matched $f_p$ against $f_i$.

Overall, this final set of results indicates that (i) the $\eta$ response of the defect modes in both multi-defect ALs are of a similar nature to that of the quarter-wave single defect grating [19]; (ii) the phase-matched $f_p$ of the band-edge modes are strictly dependent on $f_i$ (idler wave), whereas they are pinned at nearly identical values (minimal variation) for AL defect modes, even if $f_i$ of the defect mode has significantly changed. The $\eta$ sinc-plots with $gL_g \neq 0$ (Fig. 3(c1–c3) and Fig. 4(b, c)), further demonstrate that while the applied gain can alter efficiency values for all four cases by almost three orders of magnitude, unlike the uniform grating band edge modes, the AL defect modes exhibit nearly degenerate sinc-plots; it is also shown in the insets that, as expected, $\eta$ matches the laser spectrum, without invoking any further mode competition. What is critical in this context is that although the contrast between the respective highest bands and backgrounds of both $\tilde{n}_g$ and $\eta$ modulation, achieved using only $\kappa L_g$ = 3 in Fig. 2(b3 & f3) and Fig. S2(b & f), is below 5:1 at peak, due to the localized resonances, the associated cavity modes provide the basis for strongly amplified filtering, which not only leads to laser modal gain modulation but also boosts $\eta$ contrast by several orders, which is sufficient for targeted WDM applications (Fig. 1). The holographic design allows multiple THz channels with a minimum separation $\Delta f_{RF}$, which scales inversely with the grating element number [16]; a user can therefore set the splitting of SFG/DFG channels to tens of GHz (Fig. 4(a)). Finally, a more realistic conversion bandwidth can be estimated, limited only by the underlying FP cavity that provides a linear dispersion [5], leading to a THz idler bandwidth of > 200 GHz for a single pump frequency.

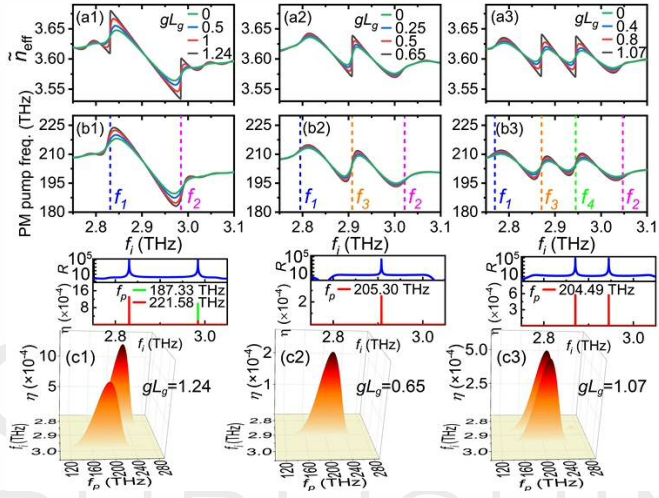

Fig. 3. Effect of gain on the nonlinear TMM simulation. (a1) – (a3) $\tilde{n}_{eff}$ against $f_i$ (dispersion) compared with (b1) – (b3) phase matched $f_p$ against $f_i$, revealing that, while the $f_p$ for the band-edge modes varies, they remain robust for the defect modes, which show no movement as $gL_g$ varies. (c1) – (c3) $\eta$ over the $f_i$ - $f_p$ plane, exhibiting that gain (corresponding to the threshold $gL_g$ values from Fig. 2 (d1 – d3) respectively)-induced changes in modal power can alter $\eta$ values by almost three orders of magnitude (note the degenerate $f_p$ values in (c3)); insets: reflection gain (power) $R$ and $\eta$ against $f_i$.

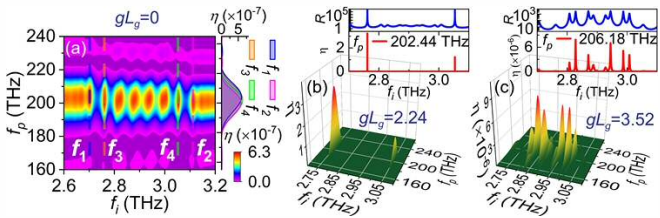

Fig. 4. $\eta$ for the multi-wavelength long AL design [21]; parameters are as in Fig. 2. (a) 3D plot ($gL_g = 0$), exhibiting degenerate $f_p$ values for the entire ensemble of defect modes; (b) & (c) $\eta$ over the $f_i$ - $f_p$ plane shown for two example $gL_g$ values (from Fig. S2(d)); insets: $R$ and $\eta$ against $f_i$.

In conclusion, we have studied the nonlinear frequency conversion process in holographically-designed AL lasers, where the lattices are designed to provide dispersion modulation at the idler band. The key findings are, firstly, both the group and effective index dispersion responses of the gratings play a significant role in controlling the nonlinear conversion process; while the conversion efficiency is strongly linked with the group index response (i.e. the photonic DOM), the effective index dispersion determines the choice of the necessary phase-matched pump frequency. Finally, although for the uniform and single-defect DFBs, the phase-matching process displays traditional distinct sinc-plots for the specific choices of the idler modes, the defect modes of the AL lasers have identical phase-matched pump frequencies; this may find potential applications in developing coherent THz WDM systems.

**Acknowledgments.** W.J. and L.H. acknowledge the University of Manchester's President's Doctoral Scholar (PDS) award.

**Author Contributions.** W.J. and L.H. developed the TMM framework. W.J. discussed the results, created figures, and contributed to Supplement preparation. S.C. conceived and supervised the project, designed the structures, interpreted the results, and wrote the manuscript.

**Supplemental document**. See Supplement 1 for supporting content.

**Disclosures.** The authors declare no conflicts of interest.

# Degenerate phase-matching for multi-wavelength nonlinear mixing in aperiodic lattice lasers: supplement

## WEI JIANG, LI HUA, SUBHASISH CHAKRABORTY*

*Department of Electrical and Electronic Engineering, University of Manchester, M13 9PL, UK*
*Corresponding author: s.chakraborty@manchester.ac.uk*

As shown in Fig. S1, the plane wave coefficients of the idler ($E_i$), pump ($E_p$), and signal ($E_s$) waves are exploited for construction of the nonlinear transfer matrix for a single element, which is essential to build the overall transfer matrix of the grating. Following a formalism similar to that for a uniform DFB grating [1], the grating element scattering strength is defined by the coupling constant $\kappa L_g$ as $\kappa = 2\Delta n/\lambda_B$, where $\Delta n$ is the refractive index contrast given by $\Delta n = n_1 - n_2$, $L_g$ is the physical length of the grating, and $\lambda_B$ is the Bragg wavelength given by $\lambda_B = c/f_B = 2\overline{n_{\text{eff}}}\Lambda$ ($\overline{n_{\text{eff}}}$ is the average refractive index, $\Lambda$ is the minimum separation between the grating elements and $c$ is the speed of light in free space). The second-order nonlinear susceptibility ($\chi^{(2)}$) of GaAs used here is 100 pm/V [2] and its refractive index in the NIR spectral region is calculated using the dispersion formula from Ref. [3]. Given that our previous experimental work (Ref. [4]) reports the use of 3 mW injected pump power (Yenista OSICS T100), accordingly, here we assume that $P_{\text{pump}} = 3$ mW; in the same paper, the cross-section of the single-metal QCL waveguide (12.8 μm-high, 200 μm-wide) is similar to that reported in Ref. [5] and so, for a better match, we assumed that the effective mode overlap is $1/A = 4\times10^8$ m$^{-2}$.

The electric field of the signal wave can be derived from the following wave equation [6, 7]:

$$\nabla^2 E_{s(m)}(z) + k_{s(m)}^2 E_{s(m)}(z) = -2k_{s(0)}^2 E_{p(m)}(z)E_{i(m)}(z). \tag{1}$$

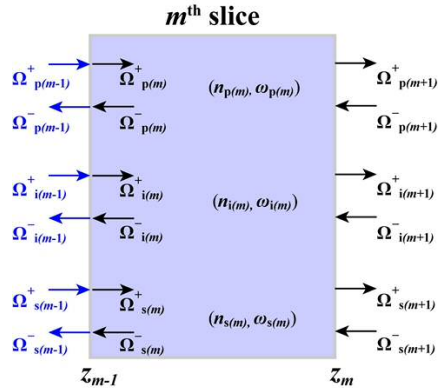

Fig. S1. Schematics of nonlinear TMM formulation involving the plane wave coefficients of pump ($E_p$), idler ($E_i$), and signal ($E_s$) fields at the boundaries of the $m^{\text{th}}$ element of a grating.

and the general solution of Eq. (1) can be expressed as:

$$E_{s(m)}(z) = \Omega_{s(m)}^{+} exp\left[ik_{s(m)}(z - z_{(m-1)})\right] + \Omega_{s(m)}^{-} exp\left[-ik_{s(m)}(z - z_{(m-1)})\right]$$
$$+ A\Omega_{p(m)}^{+}\Omega_{i(m)}^{+} exp\left[i(k_{p(m)} + k_{i(m)})(z - z_{(m-1)})\right]$$

$$+ A\Omega^-_{p(m)}\Omega^-_{i(m)} exp\left[-i(k_{p(m)} + k_{i(m)})(z - z_{(m-1)})\right]$$
$$+ C\Omega^+_{p(m)}\Omega^-_{i(m)} exp\left[i(k_{p(m)} - k_{i(m)})(z - z_{(m-1)})\right]$$
$$+ C\Omega^-_{p(m)}\Omega^+_{i(m)} exp\left[-i(k_{p(m)} - k_{i(m)})(z - z_{(m-1)})\right]. \quad (2)$$

Here, $\Omega_{j(m)}$ (j=p, i, or s) is the plane wave coefficient of the electric field of the $m^{th}$ element and $k_{j(m)} = n_{j(m)}k_{j0} = n_{j(m)}\omega_{j(m)}/c$ (where $n_{j(m)}$ is the refractive index, $\omega_{j(m)}$ is the angular frequency, and $c$ is the speed of light in a vacuum). The sign + (-) represents the forward (backward) propagating wave. The coefficients $A$ and $C$ are given as $A = -2k^2_{s(0)}\chi^{(2)}/(k^2_{s(m)} - ((k_{p(m)} + k_{i(m)})^2)$ and $C = -2k^2_{s(0)}\chi^{(2)}/(k^2_{s(m)} - ((k_{p(m)} - k_{i(m)})^2)$.

The signal wave ($E_s$) is generated by the nonlinear mixing of the $E_p$ and $E_i$ waves. Therefore, the nonlinear transfer matrix of the signal wave consists of two parts: linear component **t** (2×2) of $E_s$ and nonlinear component **r** (2×1) due to the mixing of $E_p$ and $E_i$. Based on the boundary conditions for the electric and magnetic fields on either side of the $m^{th}$ element, the nonlinear transfer matrix can be expressed as:

$$\begin{pmatrix} E^+_{s(m)} \\ E^-_{s(m)} \end{pmatrix} = t \begin{pmatrix} E^+_{s(m-1)} \\ E^-_{s(m-1)} \end{pmatrix} + r. \quad (3)$$

Using the recursion algorithm [1, 2], the overall transfer matrix for a DFB grating can be expressed as:

$$\begin{pmatrix} E^+_{st} \\ E^-_{st} \end{pmatrix} = T \begin{pmatrix} E^+_{s0} \\ E^-_{s0} \end{pmatrix} + R = \begin{pmatrix} T_{11} & T_{12} \\ T_{21} & T_{22} \end{pmatrix} \begin{pmatrix} E^+_{s0} \\ E^-_{s0} \end{pmatrix} + \begin{pmatrix} R_1 \\ R_2 \end{pmatrix}. \quad (4)$$

where $E_{st}$ and $E_{s0}$ are the amplitudes of the signal field at the output and input sides of the grating, respectively. The detailed derivations of **T** and **R** are presented in Ref. [2]. Because there is no input signal wave, $E^+_{s0}$ and $E^-_{st}$ should be 0. Consequently, the amplitude of the forward propagating signal wave $E^+_{st}$ can be derived from Eq. (4) as:

$$E^+_{st} = T_{12}(-R_2/T_{22}) + R_1. \quad (5)$$

From Eq. (5), the nonlinear conversion efficiency for the forward propagating signal wave can be written as $\eta = (E^+_{st})^2/(E^+_{p0})^2$, where $E^+_{p0}$ is the amplitude of the incident pump wave. Here, we note that TMM is primarily a steady-state analysis tool and does not inherently account for dynamic effects such as transient behaviors, modulation response, and noise characteristics. Furthermore, our TMM framework works under the principle that the phase-matching conditions for both SFG and DFG are the same, i.e. in every sense, SFG and DFG sidebands are perfectly symmetrical, and they would produce identical plots; hence, in our simulation, only SFG is considered.

The spatial distributions of the four gratings, i.e. uniform, single-defect and the two aperiodic DFB structures [8-10], are depicted in Fig. 1(a1) – (a3) and Fig. S2(a). The linear TMM is used to calculate the optical properties of these gratings, such as the power reflectivity (R) and the group delay ($\tau_g$). Detailed descriptions of the linear TMM calculations are presented in Refs. [9-12]. Briefly, $\tau_g$ is calculated based on the transmission phase ($\varphi$) of the grating as $\tau_g = d\varphi/d\omega$, which is proportional to the photonic DOM [13], as shown in Fig. 1(b1) – (b3) and Fig. S2(b). To analyze the dispersion of the grating, the group index ($\tilde{n}_g$) (Fig. 1(b1) – (b3), Fig. S2(b)) and effective index ($\tilde{n}_{eff}$) (Fig. 1(c1) – (c3), Fig. S2(c) are calculated, as $\tilde{n}_{eff} = (c/L_g)(\varphi/\omega)$ and $\tilde{n}_g = (c/L_g)(d\varphi/d\omega)$ [14]. Contour plots of the power reflectivity as a function of $f_i$ and modal gain ($gL_g$) are shown in Fig. 1(d1) – (d3) and Fig. S2(d), where a series of singularities (R → +∞) can determine the frequencies of the resonance modes and the corresponding values of the threshold gain.

For the uniform grating, the enhancement of $\tau_g$ at the band edges leads to two symmetrical band-edge modes with identical modal gain (Fig. 1(d1)), whereas for the single-defect grating, a defect mode is generated at $f_B$, together with the two band-edge modes (Fig. 1(d2)). The defect mode ($f_3 = f_B = 2.9078$ THz) has a significantly lower threshold gain than the band-edge modes ($f_1 = 2.795$ THz and $f_2 = 3.0218$ THz). The aperiodic lattice (Fig. 1(a3)), in contrast, provides two defect resonances ($f_3 = 2.8706$ THz and $f_4 = 2.945$ THz) whose threshold gains are lower than those of the outer band-edge modes ($f_1 = 2.7692$ THz and $f_2 = 3.047$ THz) (Fig. 1(d3)) [10]. As previously demonstrated in the context of frequency selection and electronic switching, these two modes are of similar nature to that of the single-defect DFB, in terms of their spectral precision as well as their insensitivity to the grating coupling strength $\kappa L_g$.

Finally, in order to further highlight the impact of an aperiodic lattice modulation, i.e. generating multiple THz channels and then up-converting them in the optical domain (SFG/DFG channels, with a high level of precision, selectivity and contrast) for WDM applications, as targeted in Fig. 1, here we introduce a new set of nonlinear TMM simulations, involving another (long) aperiodic lattice (AL) design (Fig. S2(a)), first reported in Refs. [8, 9]. The AL was designed to provide multiple defect modes with a mode separation of ~ $0.01f_B$, (29 GHz) [8]. There are 8 defect modes between 2 band-edge modes located at $f_1=2.7044$ THz and $f_2=3.1112$ THz, as shown in Fig. S2(d). As expected, the calculated nonlinear conversion efficiency $\eta$ (Fig. S2(e)) shows a strong correlation with DOM (Fig. S2(b)). For the two band-edge modes, the calculated PM pump frequencies ($f_p$) are 204.08 THz and 200.16 THz. In contrast, the $f_p$ for the two defect modes, which are separated by 289.5 GHz ($f_3=2.7641$ THz and $f_4=3.0536$ THz, which are the lowest threshold modes), are computed; however, despite the large separation, computed $f_p$ values exhibit the nearly degenerate values at 202.48 THz and 201.52 THz, respectively. As explained in the main manuscript, this is owing to the fact that $f_p$ variation against $f_i$ (Fig. S2(g)) follows a one-to-one correlation with the grating dispersion ($\tilde{n}_{eff}$ vs. $f_i$ plot, Fig. S2(c)). The robustness of the AL defect modes can be further highlighted by comparing their $\eta$ resonance positions for $gL_g \neq 0$ (Fig. S3), where the effect of gain on the nonlinear TMM simulation is presented – unlike the band edge modes, the phase matched $f_p$ for the defect modes remain robust against the gain variation.

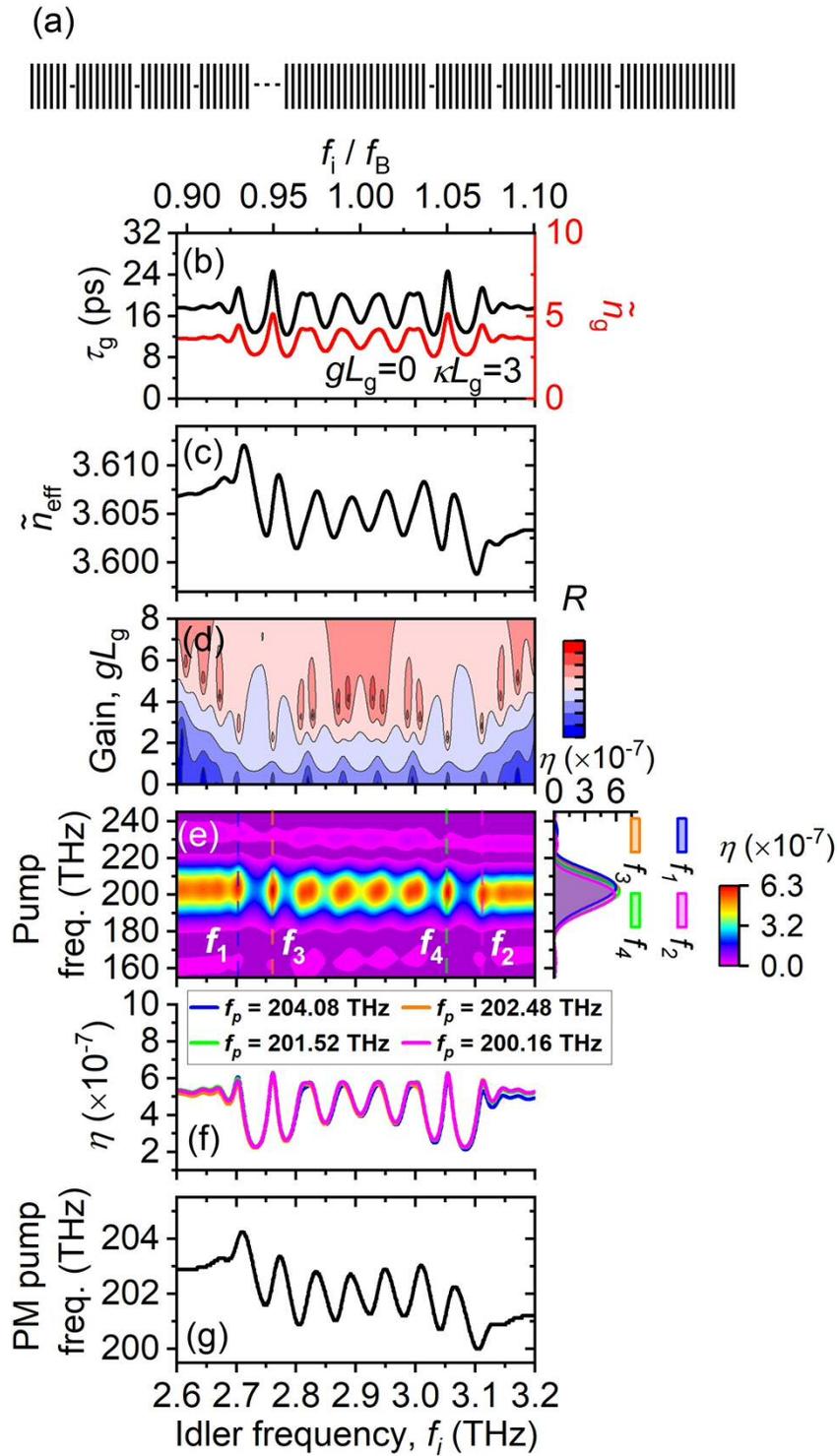

Fig. S2. Nonlinear TMM simulation results as a function of the idler frequency ($f_i$) for the long aperiodic DFB grating; the resonant responses of group delay ($\tau_g$) or group index ($\tilde{n}_g$) functions lead to the resonant enhancement of the nonlinear conversion efficiency ($\eta$) (for SFG), whereas the effective index ($\tilde{n}_{eff}$) dispersion determines the choice of the necessary phase-matched pump frequency ($f_p$). Parameters used are: $\overline{n}_{eff}$ = 3.605, $\kappa L_g$ = 3, $gL_g$ = 0, $\Lambda$ = 14.3 μm, $\chi^{(2)}$ (GaAs) = 100 pm/V, the injected pump power $P_{pump}$ = 3 mW and the effective mode overlaps $1/A$ = 4×10$^8$ m$^{-2}$. (a) Grating design – lines represent low index; horizontal dashes represent defects in multiples of $\Lambda/2$; (b) $\tau_g$ (left) and $\tilde{n}_g$ (right); (c) $\tilde{n}_{eff}$; (d) Reflection gain (power), $R$ over the $f_i$-$gL_g$ plane; (e) $\eta$ over the $f_i$–$f_p$ plane (the color 3D plots) and $\eta$ against $f_p$ (right panel); (f) $\eta$ against $f_i$; (g) Phase matched $f_p$ against $f_i$.

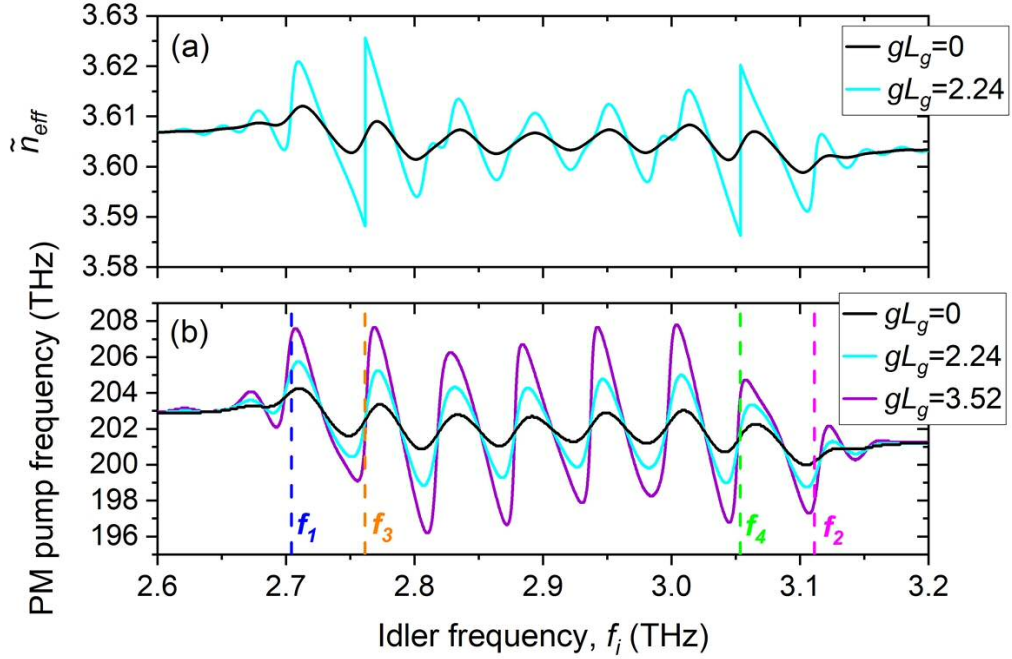

Fig. S3. Effect of gain on the nonlinear TMM simulation. (a) $\tilde{n}_{\text{eff}}$ dispersion compared with (b) phase matched $f_p$ as $gL_g$ varies, revealing that the $f_p$ values remain robust for the ten defect modes, between $f_3$ and $f_4$, which are separated by an average spacing ~ $0.01 f_B$, spanning over ~290 GHz, showing no movement as the gain is varied.